\documentclass[fleqn,usenatbib]{mnras}

\usepackage{newtxtext,newtxmath}

\usepackage[T1]{fontenc}

\DeclareRobustCommand{\VAN}[3]{#2}
\let\VANthebibliography\thebibliography
\def\thebibliography{\DeclareRobustCommand{\VAN}[3]{##3}\VANthebibliography}

\usepackage{graphicx}	
\usepackage{amsmath}	
\usepackage{amssymb}	
\usepackage{booktabs}
\usepackage{multirow}
\usepackage{adjustbox}
\usepackage{color,xcolor}
\usepackage{threeparttable}

\title[High-$z$ galaxies: JWST's seen and unseen]{Seen and unseen: bursty star formation and its implications for observations of high-redshift galaxies with JWST}

\author[G.~Sun et al.]{Guochao~Sun$^{1}$\thanks{E-mail: guochao.sun@northwestern.edu}, Claude-Andr\'{e} Faucher-Gigu\`{e}re$^{1}$, Christopher C. Hayward$^{2}$ and Xuejian Shen$^{3}$ \\
$^{1}$CIERA and Department of Physics and Astronomy, Northwestern University, 1800 Sherman Ave, Evanston, IL 60201, USA \\
$^{2}$Center for Computational Astrophysics, Flatiron Institute, 162 Fifth Avenue, New York, NY 10010, USA \\
$^{3}$TAPIR, California Institute of Technology, Pasadena, CA 91125, USA
}

\date{Accepted XXX. Received YYY; in original form ZZZ}

\pubyear{2023}

\begin{document}

\defcitealias{FZ_2013}{FZ13}

\label{firstpage}
\pagerange{\pageref{firstpage}--\pageref{lastpage}}
\maketitle

\begin{abstract}
Both observations and simulations have shown strong evidence for highly time-variable star formation in low-mass and/or high-redshift galaxies, which has important observational implications because high-redshift galaxy samples are rest-UV selected and therefore particularly sensitive to the recent star formation. Using a suite of cosmological `zoom-in' simulations at $z>5$ from the Feedback in Realistic Environments (FIRE) project, we examine the implications of bursty star formation histories for observations of high-redshift galaxies with \textit{JWST}. We characterize how the galaxy observability depends on the star formation history. We also investigate selection effects due to bursty star formation on the physical properties measured, such as the gas fraction, specific star formation rate, and metallicity. We find the observability to be highly time-dependent for galaxies near the survey's limiting flux due to the SFR variability: as the star formation rate fluctuates, the same galaxy oscillates in and out of the observable sample. The observable fraction $f_\mathrm{obs} = 50\%$ at $z \sim 7$ and $M_{\star} \sim 10^{8.5}$ to $10^{9}\,M_{\odot}$ for a \textit{JWST}/NIRCam survey reaching a limiting magnitude of $m^\mathrm{lim}_\mathrm{AB} \sim 29$--30, representative of surveys such as JADES and CEERS. \textit{JWST}-detectable galaxies near the survey limit tend to have properties characteristic of galaxies in the bursty phase: on average, they show approximately 2.5 times higher cold, dense gas fractions and 20 times higher specific star formation rates at a given stellar mass than galaxies below the rest-UV detection threshold. Our study represents a first step in quantifying selection effects and the associated biases due to bursty star formation in studying high-redshift galaxy properties. 
\end{abstract}





\section{Introduction}

The picture of when and how the universe formed its first generation of galaxies has been revolutionized since the first data release from the \textit{James Webb Space Telescope (JWST)}, even only at the dawn of the \textit{JWST} era. Not only has it enabled unprecedented measurements of the physical properties of the high-redshift galaxy population, such as their mass, star formation rate (SFR), number density, morphology, and chemical evolution \cite[][]{Curtis-Lake_2022, Ferreira_2022, Finkelstein_2022, Labbe_2022, Naidu_2022, Donnan_2023, Harikane_2023, Robertson_2023, Shapley_2023}, the race is on to answer the many intriguing questions posed by the early results \citep{Boylan-Kolchin_2022, Ferrara_2022, Kohandel_2023, Mason_2023, MF_2023}. 

The search for and identification of high-$z$ galaxy candidates by \textit{JWST} rely primarily on the rest-UV, multi-band photometry in search of the Lyman break. This so-called photometric `drop-out' technique \citep{Steidel_1996} has proven to be extremely powerful in the hunt of the most distant galaxies over the last three decades using both ground-based and space telescopes. Several major \textit{JWST} observing programs using this technique are now underway to push the observational frontier of high-$z$ galaxies to newer limits, such as CEERS \citep{Finkelstein_2023}, COSMOS-Web \citep{Casey_2022}, GLASS-JWST \citep{Treu_2022}, JADES \citep{Williams_2018,Robertson_2023NatAs}, and UNCOVER \citep{Bezanson_2022}. Given that the galaxy spectral energy distribution (SED) in the rest-UV is highly sensitive to the recent star formation history (SFH), a good understanding of the SFH is essential to not only the success of the `drop-out' technique \citep{Furlanetto_2022} but also the inference of physical information that follows \citep{Endsley_2022,Tacchella_2022}. 

While most of the SED fitting codes on the market include simple parameterization of time-variable SFHs, both observations and simulations suggest that star formation in high-$z$ galaxies may have occurred in a highly bursty manner \citep{Muratov_2015, Dominguez_2015,Guo_2016,Caputi_2017,Sparre_2017,CAFG_2018,Faisst_2019, Rinaldi_2022, Dressler_2023a, Dressler_2023b, Endsley_2023, Looser_2023bursty} not well-described by simple parameterized models, likely due to the strong effects of stellar feedback \citep{Stern_2021,Furlanetto_2022_BurstySF,Hopkins_2023}. The burstiness (or duty cycle) of star formation is therefore an important factor to consider in the interpretation of the observed galaxy samples at high redshifts. 

\begin{table}
\centering
\caption{Properties of the most massive (or the `primary') galaxy in each of the 26 simulations analyzed in this work, including the virial mass, stellar mass, cold and dense gas fraction, specific SFR, and gas-phase metallicity, evaluated at $z=6$ (for \texttt{z5mxx}), 8 (for \texttt{z7mxx}), and 10 (for \texttt{z9mxx}).}
\begin{threeparttable}
\begin{tabular}{cccccc}
\hline
Simulation & $M_\mathrm{vir}$ & $M_{\star}$ & $f_\mathrm{gas,CD}$ & $\mathrm{sSFR}$ & $Z_\mathrm{gas}$ \\
${}$ & $(M_{\odot})$ & $(M_{\odot})$ & ${}$ & $(\mathrm{yr^{-1}})$ & $(Z_{\odot})$ \\
\hline
\texttt{z5m11a} & 1.6e+10 & 4.0e+7 & 3.4e$-$3 & 7.4e$-$10 & 3.0e$-$1 \\
\texttt{z5m11b} & 2.3e+10 & 6.7e+7 & 1.4e$-$1 & 1.1e$-$9 & 2.9e$-$2 \\
\texttt{z5m11c} & 6.2e+10 & 3.6e+8 & 3.5e$-$1 & 1.6e$-$8 & 1.1e$-$1 \\
\texttt{z5m11d} & 9.5e+10 & 8.9e+8 & 2.1e$-$1 & 6.0e$-$9 & 1.4e$-$1 \\
\texttt{z5m11e} & 1.2e+11 & 3.4e+8 & 3.6e$-$1 & 9.6e$-$10 & 6.3e$-$2 \\
\texttt{z5m11f} & 1.7e+11 & 6.2e+8 & 2.8e$-$1 & 5.8e$-$9 & 1.2e$-$1 \\
\texttt{z5m11g} & 9.1e+10 & 2.1e+8 & 6.5e$-$1 & 4.4e$-$9 & 4.3e$-$2 \\
\texttt{z5m11h} & 7.8e+10 & 3.8e+8 & 5.9e$-$1 & 9.7e$-$9 & 6.2e$-$2 \\
\texttt{z5m11i} & 2.6e+10 & 7.2e+7 & 0.0e$+$0 & 1.4e$-$12 & 2.7e+0 \\
\texttt{z5m12a} & 2.3e+11 & 1.4e+9 & 2.4e$-$1 & 1.1e$-$8 & 1.5e$-$1 \\
\texttt{z5m12b} & 4.7e+11 & 1.0e+10 & 1.2e$-$1 & 9.2e$-$9 & 2.4e$-$1 \\
\texttt{z5m12c} & 4.7e+11 & 2.7e+9 & 3.2e$-$1 & 2.2e$-$9 & 1.6e$-$1 \\
\texttt{z5m12d} & 3.3e+11 & 1.9e+9 & 3.6e$-$1 & 1.3e$-$8 & 1.2e$-$1 \\
\texttt{z5m12e} & 3.1e+11 & 4.3e+9 & 2.3e$-$1 & 1.9e$-$8 & 2.0e$-$1 \\
\texttt{z7m11a} & 2.3e+11 & 1.8e+9 & 4.3e$-$1 & 3.7e$-$8 & 1.0e$-$1 \\
\texttt{z7m11b} & 5.2e+10 & 5.5e+7 & 7.3e$-$1 & 1.6e$-$8 & 3.3e$-$2 \\
\texttt{z7m11c} & 1.1e+11 & 8.0e+8 & 4.5e$-$1 & 2.0e$-$8 & 7.2e$-$2 \\
\texttt{z7m12a} & 2.9e+11 & 2.6e+9 & 2.3e$-$1 & 8.6e$-$9 & 1.6e$-$1 \\
\texttt{z7m12b} & 3.3e+11 & 5.8e+9 & 1.5e$-$1 & 9.5e$-$9 & 1.9e$-$1 \\
\texttt{z7m12c} & 2.7e+11 & 1.9e+9 & 3.0e$-$1 & 8.2e$-$9 & 1.2e$-$1 \\
\texttt{z9m11a} & 9.1e+10 & 2.9e+8 & 4.8e$-$1 & 2.0e$-$8 & 6.7e$-$2 \\
\texttt{z9m11b} & 1.2e+11 & 1.3e+9 & 2.2e$-$2 & 4.7e$-$9 & 3.2e$-$1 \\
\texttt{z9m11c} & 1.1e+11 & 1.3e+9 & 1.5e$-$1 & 1.8e$-$8 & 2.0e$-$1 \\
\texttt{z9m11d} & 6.8e+10 & 2.9e+8 & 1.5e$-$3 & 6.9e$-$10 & 3.8e$-$1 \\
\texttt{z9m11e} & 7.8e+10 & 3.6e+8 & 2.3e$-$1 & 1.5e$-$8 & 8.1e$-$2 \\
\texttt{z9m12a} & 3.1e+11 & 6.6e+9 & 1.8e$-$1 & 1.3e$-$8 & 1.8e$-$1 \\
\hline
\end{tabular}
\begin{tablenotes}
\item \textit{Note}: Values of $f_\mathrm{gas,CD}$, sSFR, and $Z_\mathrm{gas}$ can vary strongly in time as the SFR fluctuates.
\end{tablenotes}
\end{threeparttable}
\label{tb:gal_properties}
\end{table}

In this paper, using simulations from the Feedback in Realistic Environments (FIRE) project\footnote{See the FIRE project website: \url{http://fire.northwestern.edu}.}, we seek to answer the following questions about bursty star formation and its implications for high-$z$ galaxy observations with \textit{JWST}: How does the SFR variability impact the galaxy observability by \textit{JWST}? Are there any selection effects on inferred galaxy properties caused by bursty star formation? Answers to these questions not only provide key insights to the interpretation of existing and forthcoming \textit{JWST} observations, but also motivate further studies on the role of bursty star formation in early galaxy formation. In \citet{Sun_2023}, for example, we show that these are the same simulations that match the UV luminosity functions measured by \textit{JWST} at $8 \lesssim z \lesssim 12$, of which the bright end is substantially affected by bursty star formation. Throughout this work, we adopt a flat $\Lambda$CDM cosmology consistent with \citet{Planck_2016}. All magnitudes are quoted in the AB system \citep{OG_1983}. 

\section{Methods} \label{sec:methods}

\subsection{The simulations} \label{sec:methods:simulations}

We analyze the \textit{High-Redshift} suite of the FIRE-2 cosmological zoom-in simulations that were first presented in \citet{Ma_2018_size, Ma_2018_lf}. These simulations were generated with an identical version of \textsc{gizmo} \citep{Hopkins_2015} in the meshless finite-mass (MFM) mode for hydrodynamics. Physical models of the multiphase interstellar medium (ISM), star formation, and stellar feedback used in the FIRE-2 simulations are described in detail by \citet{Hopkins_2018}. A uniform, redshift-dependent ionizing background is assumed and reionizes the Universe at $z\approx10$.\footnote{The version of the ionizing background that reionizes the Universe at $z\approx10$ is a slight update of the original \cite{CAFG_2009} model from December 2011; see \url{http://galaxies.northwestern.edu/uvb-fg09}.} Several previous studies from the pre-\textit{JWST} era demonstrated that FIRE simulations based on these models reproduced the observed UV luminosity function (UVLF) and galaxy scaling relations, such as stellar mass--halo mass 
relation, the SFR--stellar mass relation, and the mass-metallicity relation (MZR), reasonably well up to $z>5$ \citep{Ma_2016,Ma_2018_lf,Ma_2019}. 

Throughout our analysis, we focus on the well-resolved, central galaxies hosted by main haloes (rather than subhaloes) in the zoom-in region of each simulation. For each central galaxy, we examine 56 consecutive snapshots saved for each simulation, spanning a redshift range of $5<z<12$ approximately every 10 to 20\,Myr apart. To analyze the physical properties of galaxies in the 26 simulations considered (Table~\ref{tb:gal_properties}), we first identify the star and gas particles in central galaxies hosted by main haloes, which are found using the spherical overdensity-based \textsc{Amiga} Halo Finder (AHF; \citealt{KnollmannKnebe_2009}). AHF locates the haloes in the simulation volume and calculates their basic information such as the centre of mass, virial mass ($M_\mathrm{vir}$), and virial radius ($R_\mathrm{vir}$). The evolving virial overdensity definition from \citet{BN_1998} is used. The stellar mass ($M_{\star}$) and gas mass ($M_\mathrm{gas}$) contents of the galaxy are then defined to be the summed masses of star and gas particles within 0.2\,$R_\mathrm{vir}$ of the halo centre, respectively. The choice of 0.2\,$R_\mathrm{vir}$ excludes most satellites and is a simple approximation for properly estimating the stellar and gas properties of central galaxies. 

\begin{figure*}
    \centering
	\includegraphics[width=0.32\textwidth]{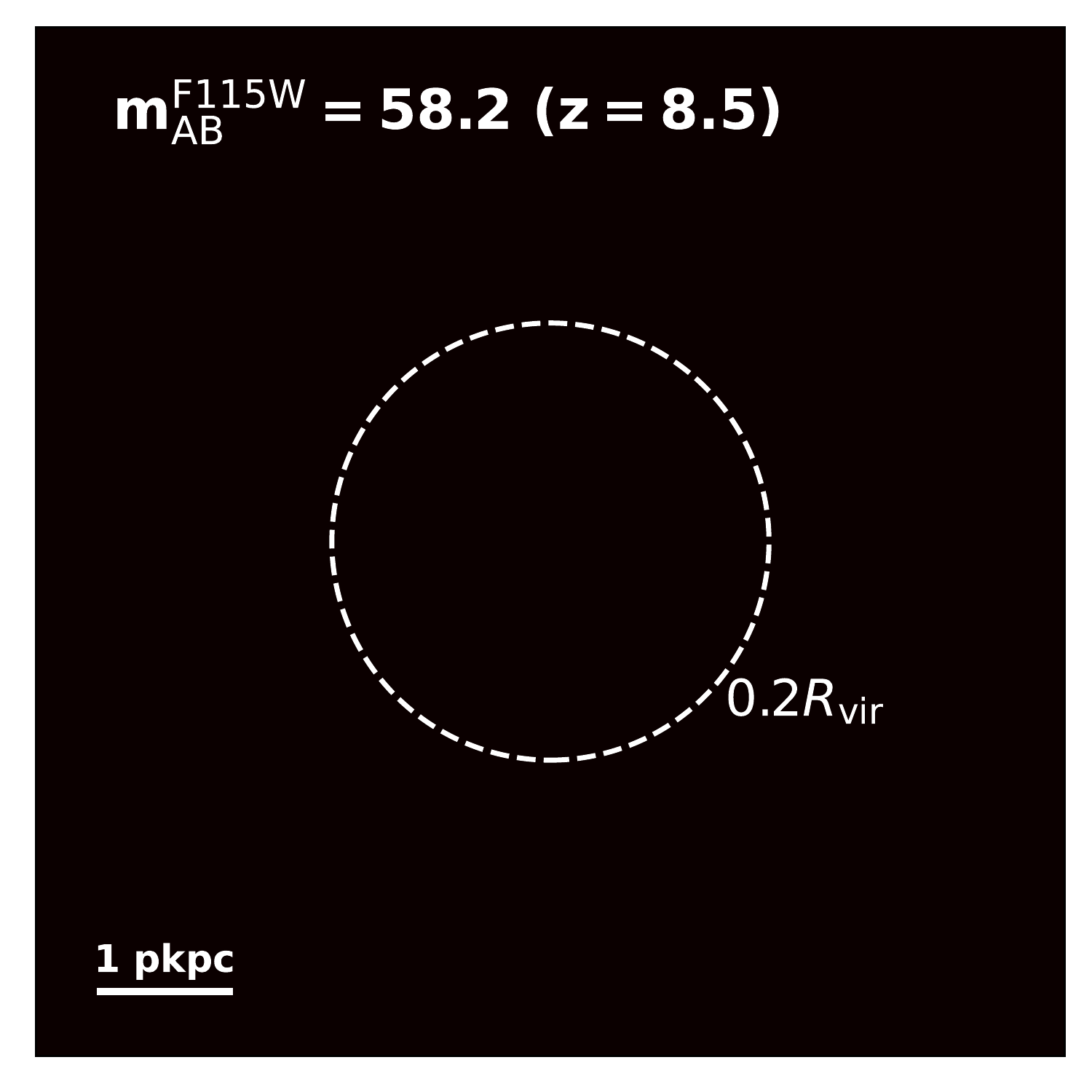}
	\includegraphics[width=0.32\textwidth]{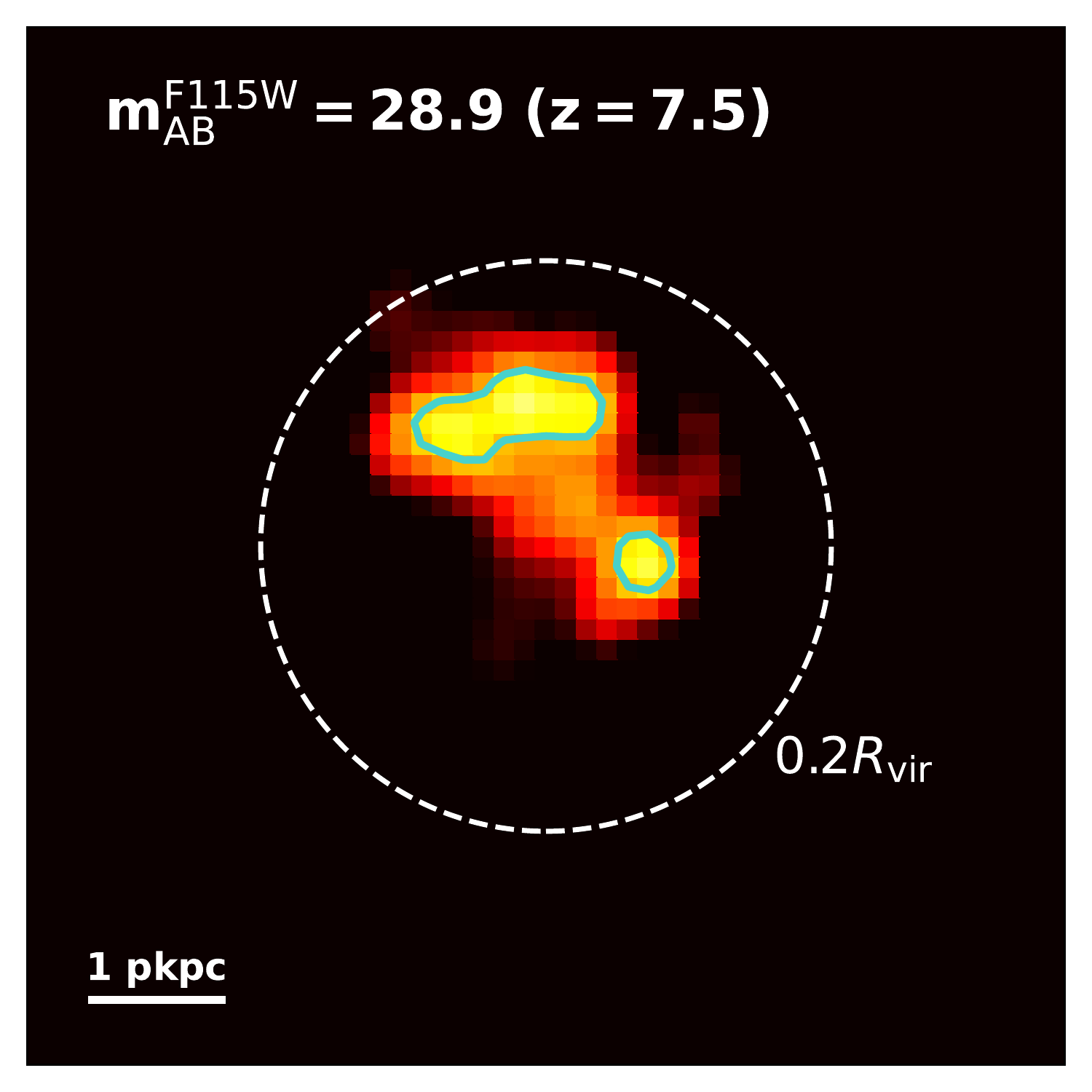}
	\includegraphics[width=0.32\textwidth]{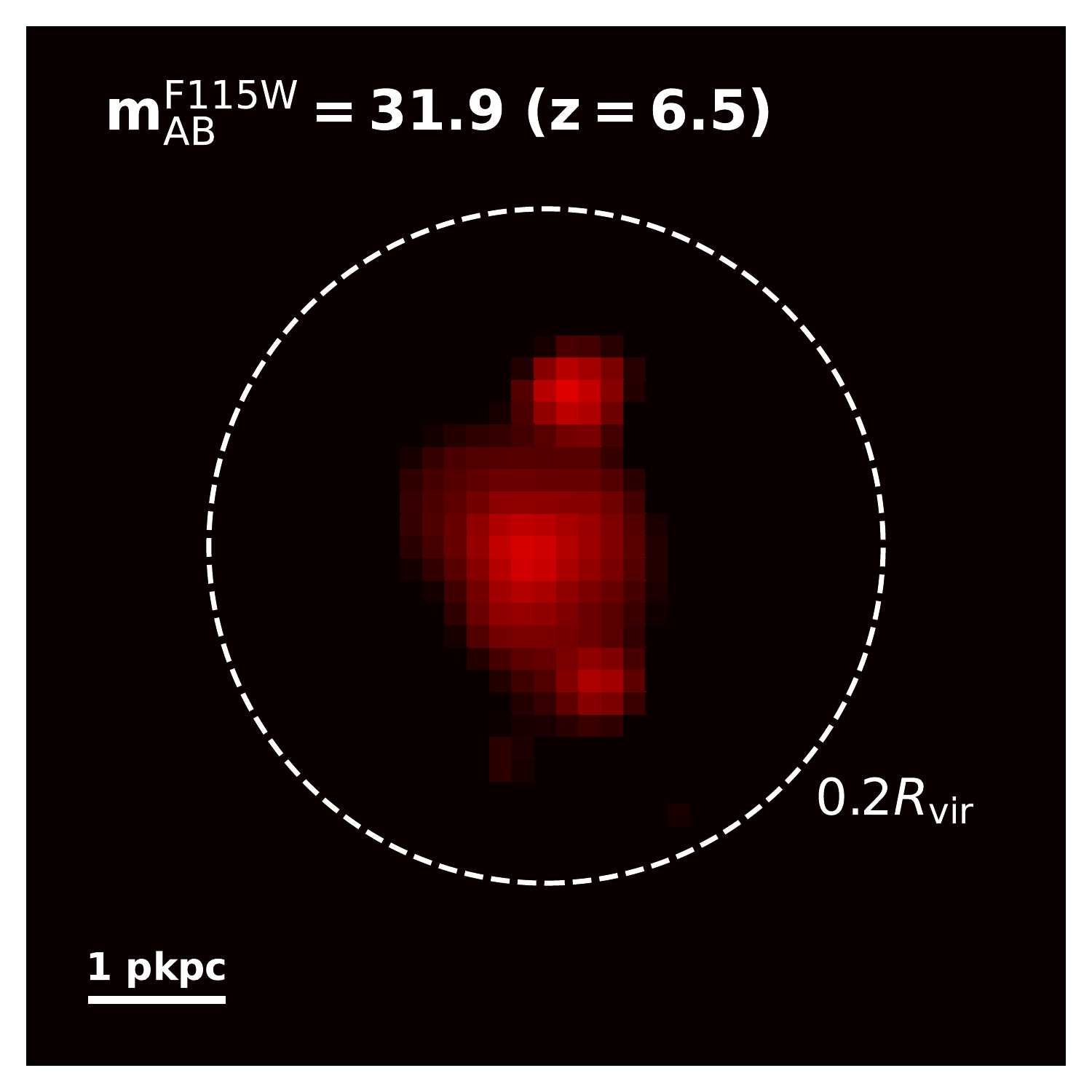}
    \caption{Noiseless, PSF-convolved \textit{JWST}/NIRCam mock images of the example galaxy \texttt{z5m11d} at $z=8.5$, 7.5, and 6.5 in band F115W with pixel size 0.031$\arcsec$. The PSF is assumed to be Gaussian, with an FWHM of 2 pixels \citep{Ma_2018_lf}. The contour marks the 1$\sigma$ surface brightness sensitivity level corresponding to $26.5\,\mathrm{mag\,arcsec^{-2}}$, which is converted from the point-source limiting magnitude of a JADES-Deep-like survey. The apparent magnitude, $m^\mathrm{F115W}_\mathrm{AB}$, of the galaxy at each redshift is labeled on the top left corner. Note the huge $m^\mathrm{F115W}_\mathrm{AB}$ value caused by strong IGM absorption when band F115W probes blueward of the Lyman break at $z\gtrsim8.5$ and the time variations of the galaxy flux and apparent size as a strong starburst occurs at $z \approx 7.5$ (see the upper panel of Fig.~\ref{fig:observability}), which causes the galaxy to oscillate into (out of) the observable regime from $z=8.5$ (7.5) to $z=7.5$ (6.5).}
    \label{fig:mock_images}
\end{figure*}

Besides the total stellar and gas masses, we measure and monitor a few other galaxy properties from the simulations to understand their connection with the duty cycle corresponding to the strongly time-variable SFH and implications this might have for \textit{JWST} observations. Using the `archaeological' approach commonly adopted for studies of the SFR variability \cite[see e.g.][]{JVF_2021,Gurvich_2023}, we estimate the instantaneous SFR from the distribution of formation times and masses of star particles evaluated at the snapshot corresponding to the redshift of interest. In the low-mass  limit most pertinent to our investigation, we expect that the archaeological approach is not significantly affected by ex-situ star formation contributed by mergers \citep[e.g.][]{Fitts_2018}. 
We have verified this to be a valid approximation for our analysis by comparing the archaeological method with the SFR estimated by evaluating the stellar mass difference between consecutive snapshots. For the gas mass fraction, we define it to be the ratio of the cold and dense gas mass to the sum of stellar and cold and dense gas masses, namely $f_\mathrm{gas,CD} = M_\mathrm{gas,CD} / (M_\mathrm{gas,CD} + M_{\star})$. For simplicity, we compute $M_\mathrm{gas,CD}$ by summing up the masses of gas particles with temperature $T < 300\,$K and density $n > 10\,\mathrm{cm^{-3}}$. These simple cuts for the gas fraction are adopted because they were found to provide a good proxy for the cold and dense molecular gas available for fueling star formation in the FIRE simulations \citep{Orr_2018}, though we note that more physically-motivated descriptions of the molecular gas content could be applied \cite[e.g.][]{Krumholz_2008,Krumholz_2009,McKee2010}. More accurate treatments of ISM chemistry in the cosmic dawn regime should be explored for detailed comparison with observations. We also consider a basic measure of the metallicity of simulated galaxies by keeping track of the overall gas-phase metallicity $Z_\mathrm{gas}=M_{Z}/M_\mathrm{gas}$, based on the mass fraction of \textit{all metals} for the gas particles. 

\subsection{Processing and analysis}

\begin{figure}
    \centering
	\includegraphics[width=\columnwidth]{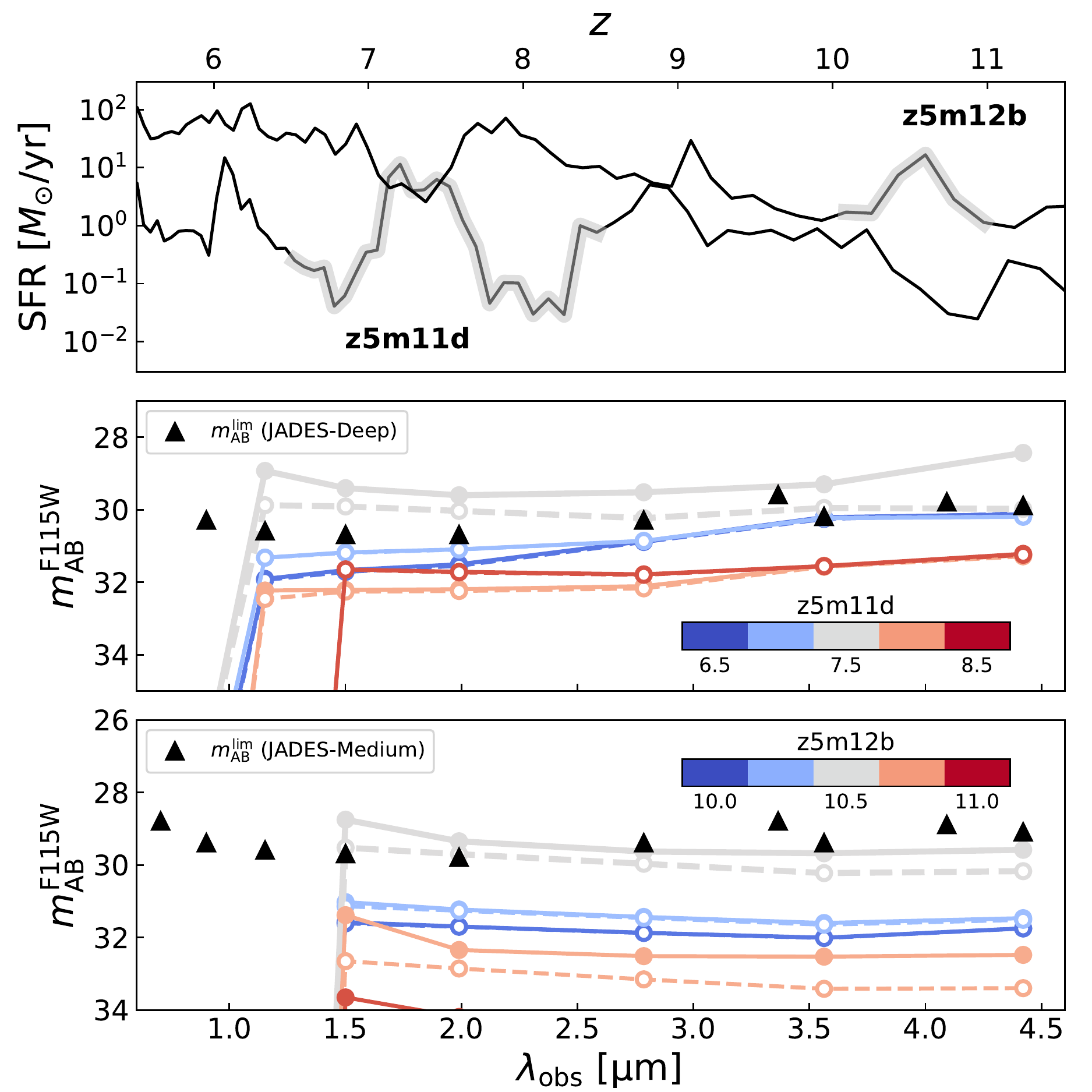}
    \caption{Impact of bursty SFH on the observability of two example galaxies, \texttt{z5m11d} and \texttt{z5m12b} from our simulations. \textit{Top:} Time variability of the SFR evaluated for an averaging timescale of 10\,Myr. Highlighted segments show the redshift intervals inspected below for the observability. \textit{Middle:} Apparent magnitudes of \texttt{z5m11d} in JWST/NIRCam filters (F070W, F090W, F115W, F150W, F200W, F277W, F356W, F444W), with (filled markers/solid lines) or without (empty markers/dashed lines) including the nebular emission. Lines are colour-coded by redshift for five snapshots of the galaxy. The triangles indicate 5$\sigma$ limiting magnitudes of the JADES-Deep survey \citep{Williams_2018}. \textit{Bottom:} Same as the middle panel but for \texttt{z5m12b} and the JADES-Medium survey. }
    \label{fig:observability}
\end{figure}

\subsubsection{Stellar and nebular emission}

To assess the observability of our simulated galaxies by \textit{JWST} surveys, we need to generate mock observations of them and compare with the survey depths. We use a modeling approach similar to \citet{Ma_2018_size,Ma_2018_lf} where synthetic, single stellar population (SSP) spectra are assigned to star particles so as to compute the apparent magnitude and therefore the detectability of each simulated galaxy. Given our interest in comparing against \textit{JWST} surveys of high-$z$ galaxies using the photometric, drop-out technique \citep{Steidel_1996,Robertson_2022}, we take into account two additional factors that are neglected in \citet{Ma_2018_size,Ma_2018_lf}, namely the nebular emission and attenuation by the neutral IGM. The IGM attenuation causes the Lyman break feature, which is the basis of the drop-out technique. The intrinsic emission spectra of galaxies are further reddened by a simple empirical prescription for dust attenuation. 

Specifically, we interpolate spectra of binary stars from BPASS v2.2 \citep{StanwayEldridge_2018} to model the rest-frame UV/optical emission spectrum of galaxies with realistic SFHs extracted from our high-$z$ simulations, which would be observed at near-infrared wavelengths by \textit{JWST}. We adopt the default stellar initial mass function (IMF), which is described by a broken power law across $0.1\,M_{\odot}<m<0.5\,M_{\odot}$ and $0.5\,M_{\odot}<m<300\,M_{\odot}$, with low-mass and high-mass slopes $\alpha_1=-1.3$ and $\alpha_2=-2.35$, respectively \citep{Kroupa_2001}. It is noteworthy that a non-universal IMF that depends on star-forming environments (e.g. metallicity, interstellar radiation field, gas temperature) could also induce stochasticity in the expected galaxy SED \citep{Chruslinska_2020, Steinhardt_2023}. However, we expect the UV variability of galaxies to be primarily driven by the time-variable SFH, given the order-of-magnitude fluctuations in the SFR, and thus postpone the investigation of such IMF effects to further work. For each star particle, we take the combination of the stellar age, $t_\mathrm{age}$, and metallicity, $Z_{\star}$, to evaluate the spectral emissivity (per unit mass of stars formed) required to scale the star particle's luminosity by its mass. Namely, the specific luminosity of the galaxy at a cosmic time $t$ can be written as $\sum_i L^{i}_{\nu}\left(t^{i}_\mathrm{age}, Z^{i}_{\star}\right)$, where $t^{i}_\mathrm{age} = t - \mathrm{SFT}_i$ is the stellar age of star particle $i$ defined as the difference between the cosmic time at the rest-frame of the observed galaxy and the star formation time of the star particle. 

Regarding nebular emission lines, although one might expect these lines to have only secondary effects on the photometry of galaxies compared to the stellar continuum, especially for wide-band filters of \textit{JWST} considered in this work, it is important to include them for better understanding e.g. their impact on the apparent magnitudes relative to effects of the SFR variability and the dependence on assumptions made for stellar populations, such as single vs. binary stars. Similar considerations also apply to the nebular continuum emission, whose contribution to the net rest-UV/optical flux can be non-negligible especially at longer wavelengths and for young stellar populations \citep{Byler_2017}. To implement the nebular emission (both lines and continuum), we use the BPASS data products from the BPASS v2.2 data release (\citealt{StanwayEldridge_2018}; see also \citealt{Xiao_2018} for an early release of nebular line predictions based on BPASS v2.1 models), where SED templates with nebular emission included are generated with \textsc{cloudy} \citep{Ferland_2017} and tabulated for grids of $t_\mathrm{age}$ and $Z_{\star}$, together with ISM properties including the gas density $n_\mathrm{H}$ and ionization parameter $U$. Assuming $n_\mathrm{H}=100\,\mathrm{cm^{-3}}$ and $U=0.01$, broadly consistent with the ionized gas properties inferred for high-$z$ galaxies from recent \textit{JWST} observations \citep{Reddy_2023}, we use the \textsc{cloudy}-processed SED templates to derive the flux contribution from each star particle. We stress that these physically motivated but ultimately simplistic prescriptions for the nebular emission are not intended to create high-accuracy mock galaxy spectra, but rather to help gauge the impact of accounting for the nebular emission in general. 

\subsubsection{IGM attenuation} \label{sec:igm}

Due to the opacity of intergalactic neutral hydrogen to UV photons, the spectra of high-$z$ galaxies are filtered by a blanketing effect causing increasing absorption blueward of Ly$\alpha$ (1216\,\AA) till the Lyman-limit wavelength (912\,\AA), beyond which the absorption becomes complete. Such IGM transmission effects create the Lyman-break feature in the observed galaxy spectrum, and we account for them with an analytic model introduced by \citet{Inoue_2014} widely used for the modeling of high-$z$ galaxy spectra recently observed by \textit{JWST}. 

\begin{figure*}
    \centering
	\includegraphics[width=\textwidth]{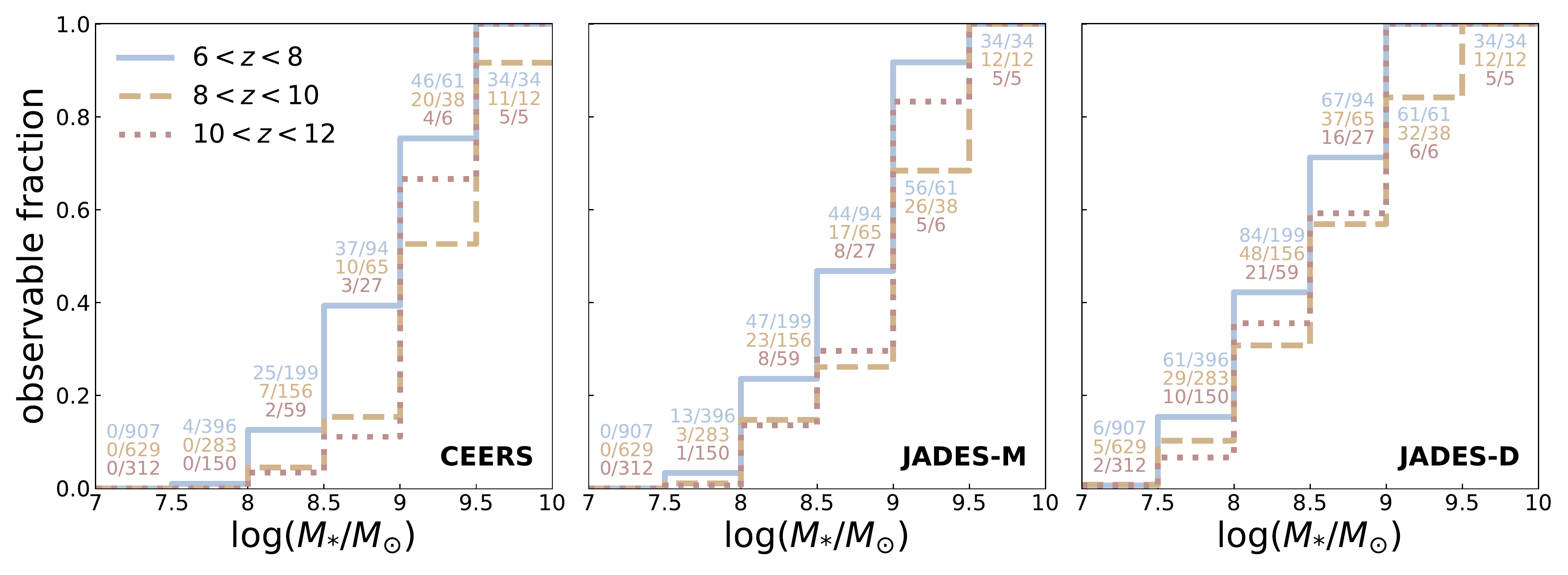}
    \caption{The observable fraction, $f_\mathrm{obs}=N_\mathrm{obs}/N_\mathrm{tot}$, defined as the ratio of detectable and total numbers of galaxies in each bin, of galaxies at $z=6$ to 12 for the three example \textit{JWST} survey depths considered (left: CEERS; middle: JADES-Medium; right: JADES-Deep). Bursty star formation affects the observability the most at mass scales where $f_\mathrm{obs} \approx 50\%$, which increase with decreasing depth or increasing redshift and roughly correspond to $M_{\star} \sim 10^{8.5}$ to $10^{9}\,M_{\odot}$ for the surveys and redshift ranges displayed.}
    \label{fig:obs_frac}
\end{figure*}

\subsubsection{Dust attenuation} \label{sec:dust}

Overall, based on empirical models for the dependence of dust attenuation on the dust content of galaxies, we expect dust attenuation to be small (i.e., $A_{\lambda} = 1.086 \tau_{\lambda} \ll 1$) for the low-mass ($M_{\star} \lesssim 10^{8.5}\,M_{\odot}$) systems at high-$z$, for which we find the largest duty cycle effects of interest to this work (see Section~\ref{sec:results}). This is further supported by several recent characterizations of dust attenuation at high redshifts based on \textit{JWST} measurements of the Balmer decrement \citep{Shapley_2023} or fits to the observed galaxy UVLFs and colours \citep{MF_2023}. For completeness, though, we include a conservative, empirical description of dust attenuation following \citet{McLure_2018}. Combining deep imaging data from the Atacama Large Millimeter Array (ALMA) at 1.3-mm and other multiwavelength data available at optical/IR of the Hubble Ultra Deep Field (HUDF), \citet{McLure_2018} study the relationship among the infrared excess, UV spectral slope, and $M_{\star}$ for a sizeable sample of star-forming galaxies at $2<z<3$ and use it to constrain the amount of dust attenuation as a function of $M_{\star}$. We adopt their best-fit relation between $A_{1600}$ and $M_{\star}$ given by a third-order polynomial in $X = \log(M_{\star}/10^{10}\,M_{\odot})$, and use it to determine $A_{\lambda}$ as a function of $M_{\star}$, assuming $\tau_{\lambda} \propto \lambda^{-1}$ as expected for a Small Magellanic Cloud-like extinction curve \citep{WeingartnerDraine_2001,Ma_2019}. 

\section{Results} \label{sec:results}

Comparing the set of post-processed simulations with \textit{JWST} observations (Section~\ref{sec:methods}), we examine how the time-variable SFH of individual high-$z$ galaxies impacts their detectability by \textit{JWST} and characterize physical properties of the galaxy population revealed or concealed by a given \textit{JWST} photometric survey. For the main results that follow, we consider the comparison with a photometric galaxy survey with depth similar to JADES-Deep \citep{Williams_2018}\footnote{JADES-Medium/Deep also utilizes two shallower mid-band filters (Figure~\ref{fig:observability}), but for simplicity we only focus on the wide-band filters.}, reaching a limiting magnitude of $m^\mathrm{lim}_\mathrm{AB} \sim 30.6$ for 5$\sigma$ point-source detections in the NIRCam wide-band filters. Besides this fiducial survey, we also inspect how these results depend on the survey depth by repeating the same analysis for two additional surveys similar to JADES-Medium and CEERS, which are approximately 1 and 2 magnitudes shallower than JADES-Deep, respectively. 

\subsection{Effects of SFR variations on the galaxy observability}

Due to the fact that the rest-UV selection for high-$z$ galaxies is highly sensitive to the recent SFH, we find a strong coherence between the SFR and apparent magnitude of the galaxies. Such a coherence leads to the migration of a given galaxy into, and out of, the observable regime as the galaxy's SFR varies, which is particularly true for galaxies near the detection threshold of the survey. Figure~\ref{fig:mock_images} shows \textit{JWST}/NIRCam mock images of the primary galaxy of \texttt{z5m11d} in band F115W at $z=8.5$, 7.5, and 6.5 (from left to right), generated after convolving high-resolution simulated images with a Gaussian point spread function (PSF) corresponding to band F115W. From these mock images, two key features of the connection between SFR variations and the observability are apparent: as the SFR of galaxies like \texttt{z5m11d} undergoes strong variations, there are both up-and-down movements in flux, as shown by the colour coding, and fluctuations of the apparent size (above the surface brightness detection limit) concurrent with flux variations, as indicated by the contours. We emphasize that these mock images are shown only for illustration purpose. For the analysis that follows, we determine the observability of galaxies by comparing the apparent magnitude against the 5$\sigma$ limiting magnitude of the survey of interest.

In Figure~\ref{fig:observability}, we illustrate the modulation of the galaxy's detectability by \textit{JWST} due to the strong time-variability of the SFR. As highlighted in the top panel, the two example galaxies, \texttt{z5m11d} and \texttt{z5m12b}, experience a major outburst of star formation at $z\sim7.5$ and $z\sim10.5$, respectively. The rapid and substantial rise and fall of the SFR associated with the starburst can temporarily brighten these originally too-faint-to-detect galaxies by several magnitudes, making them detectable for a period of time before fading off again as the massive, young stars formed in the starburst die. Throughout, we consider a galaxy to be detectable if and only if there are (1) at least two $>5\sigma$ detections redward of the Lyman break and (2) no detections blueward of it (i.e., the `drop-out' can be detected at high significance). Also noteworthy from the comparison between cases with and without the nebular emission is that changes in the observability induced by bursty star formation become more significant when the nebular emission tracing recent star formation is taken into account. 

While only two examples are shown in Figure~\ref{fig:observability}, the non-monotonic evolution of a galaxy's observability imprinted by its bursty SFH is ubiquitous in our simulations, especially in the mass and redshift range where galaxies are marginally detectable/non-detectable. To demonstrate and identify in what mass and redshift ranges galaxies are near the detection threshold and thus subject to a strong influence by bursty star formation, we show in Figure~\ref{fig:obs_frac} the observable fraction, $f_\mathrm{obs}$, together with the raw numbers of snapshots ($N_\mathrm{obs}/N_\mathrm{tot}$) used to evaluate $f_\mathrm{obs}$, of galaxies in different $M_{\star}$ bins in three redshift ranges and for three example survey depths. Overall, $f_\mathrm{obs}$ increases towards higher masses in a given redshift range, whereas at a given mass $f_\mathrm{obs}$ declines with increasing redshift. For the survey depths considered, across the redshift ranges examined, $f_\mathrm{obs} \sim 50\%$ for galaxies with $M_{\star} \sim 10^{8.5}$ to $10^{9}\,M_{\odot}$. 
This can be viewed as a characteristic mass at which bursty star formation affects observability the most. The $f_\mathrm{obs} \sim 50\%$ mass scale increases for a shallower survey (e.g. CEERS) and decreases for a deeper survey (e.g. JADES-Deep), but broadly falls within the range of $10^{8.5} \lesssim M_{\star}/M_{\odot} \lesssim 10^{9}$ for the surveys considered. For a given survey, the mass scale increases slightly with redshift. We note that other galaxy properties and their evolution with mass and redshift may also impact their observability. 

\subsection{Global properties of galaxies seen and unseen by \textit{JWST}}

Given the strong oscillations in the observability of a galaxy that are induced by its SFR variability, a natural question is whether such a modulation introduces significant selection effects. To address this question, we perform a simple classification of our simulated galaxies, after combining all the available snapshots to increase the sample size, according to their observability by our fiducial example survey (JADES-Deep). We then examine several global properties of the detectable (D) and non-detectable (ND) galaxies separately, including $f_\mathrm{gas,CD}$, the specific star formation rate (sSFR), and $Z_\mathrm{gas}$. Motivated by the distribution of $f_\mathrm{obs}$ shown in Figure~\ref{fig:obs_frac}, we focus on the fiducial survey (JADES-Deep) and consider a narrow mass bin centred around $M_{\star}=10^{8.5}\,M_{\odot}$ to minimize the influence on our analysis of mass dependence of these properties. 

\begin{figure*}
	\includegraphics[width=\textwidth]{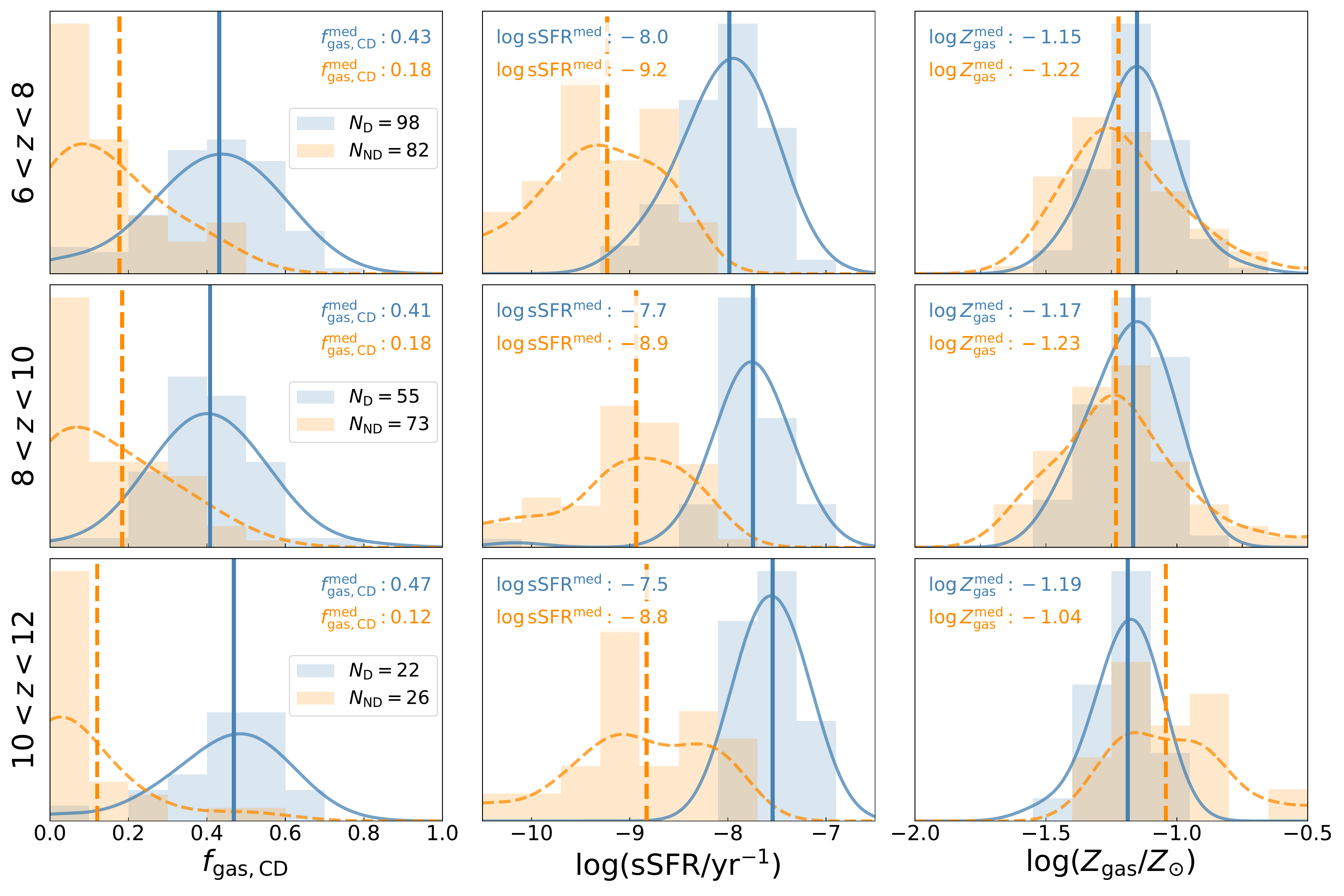}
    \caption{Distributions of physical properties ($f_\mathrm{gas}$, sSFR, $Z_\mathrm{gas}$) of galaxies at $6<z<8$ (top), $8<z<10$ (middle), and $10<z<12$ (bottom) detectable (`D', blue) and non-detectable (`ND', orange) by a JADES-Deep-like survey with \textit{JWST}. The narrow stellar mass interval, $8.2<\log(M_{\star}/M_{\odot})<8.8$, represents a mass regime where there is a mixture of detectable and non-detectable galaxies and therefore close to the detectable threshold of the survey. Median values of each property measured from a Gaussian kernel density estimation are specified on the panels for the detectable vs. non-detectable samples.}
    \label{fig:properties}
\end{figure*}

For a narrow mass range of $8.2<\log(M_{\star}/M_{\odot})<8.8$ where there is a mixture of detectable and non-detectable galaxies at roughly the same $M_{\star}$, Figure~\ref{fig:properties} shows the distributions of the three physical properties of interest estimated from our galaxies classified by their observability. As indicated by both the histograms and Gaussian kernel density estimations (KDEs), different levels of offset between the detectable and non-detectable populations are observed, which we quantify using the median value of either sample in the narrow mass bin considered. 

For $f_\mathrm{gas}$ and sSFR, clear systematic offsets exist, which suggest that detectable galaxies have preferentially more abundant cold and dense gas, as well as higher specific star formation rates. This is qualitatively in line with our expectations for galaxies having an ongoing burst of star formation --- a large mass of cold, dense molecular gas is present as fuel for star formation while many young stars are actively forming. Therefore, it is likely that galaxies being observable thanks to an ongoing starburst can lead to a significant selection bias for higher $f_\mathrm{gas}$ and sSFR, characteristic of galaxies in the starburst phase. Based on the galaxy samples displayed in Figure~\ref{fig:properties}, we estimate the systematic difference in the median $f_\mathrm{gas,CD}$ and sSFR of the detectable and non-detectable populations to be $\Delta f_\mathrm{gas,CD} \approx 0.4$\,dex and $\Delta \mathrm{sSFR} \approx 1.2$\,dex. In terms of the offset between median properties of observable galaxies and \textit{all} galaxies in the mass bin considered, we find the fractional differences to be $|f^\mathrm{obs}_\mathrm{gas,CD}-f^\mathrm{all}_\mathrm{gas,CD}|/f^\mathrm{all}_\mathrm{gas,CD} \approx 40\%$ and $|\mathrm{sSFR}^\mathrm{obs}-\mathrm{sSFR}^\mathrm{all}|/\mathrm{sSFR}^\mathrm{all} \approx 300\%$, respectively. These values reflect how representative the observable sample might be for quantifying the relation between median galaxy properties and $M_*$. 

On the other hand, we do not observe an as clear offset for $Z_\mathrm{gas}$, which arguably relates to the SFR variability less directly. Nevertheless, we note that a lower $Z_\mathrm{gas}$ for the detectable population (e.g. at $z=10$--12) could be associated with the plausible scenario where bursty galaxies `up-scattered' into the observable regime tend to have a higher supply of pristine, star-forming gas that reduces $Z_\mathrm{gas}$. While such a picture is essentially what the gas-regulator model \citep{Lilly_2013} suggests as a potential explanation for the so-called fundamental metallicity relation (FMR), more detailed analysis of the simulations is required to reach a conclusive answer, which we postpone to future work. 

In addition to the distributions shown in Figure~\ref{fig:properties} for a narrow mass bin near the detection threshold, we are also interested in quantifying how much the burstiness-induced selection effect could bias the measured galaxy properties as far as a real galaxy population is concerned. To do this, we calculate the average of each galaxy property, weighted by the halo mass function (HMF)\footnote{We compute the HMF using the \textsc{hmf} code \citep{Murray_2013}, assuming the fitting function from \citet{Behroozi_2013} optimized for high redshifts.}, $\mathrm d n / \mathrm d M_\mathrm{vir}$, evaluated at the galaxy's halo mass and redshift, for \textit{all} galaxies with $M_{\star}>10^{8.2}\,M_{\odot}$ in the detectable or non-detectable sample for the same redshift ranges shown in Figure~\ref{fig:properties}. As expected, since lower-mass galaxies are more numerous, similar offsets ($\sim0.4$\,dex for $f_\mathrm{gas,CD}$, $\sim1.2$\,dex for sSFR, and insignificant for $Z_\mathrm{gas}$) are observed between the HMF-weighted average properties of detectable and non-detectable galaxies as when considering a narrow mass bin near the detection threshold. Therefore, the offsets in galaxy properties shown in Figure~\ref{fig:properties} are broadly representative of biases expected when galaxy properties are averaged over an entire survey.

Finally, by repeating the same analysis for the other two example survey depths (JADES-Medium and CEERS), we find comparable offsets as for the fiducial case of a JADES-Deep-like depth. The main difference is that the sample size of the detectables and the mass range where the bias will have greatest impact ($f_{\rm obs}\sim 50\%$) depend on survey depth (see Figure~\ref{fig:obs_frac}). 

\section{Conclusions}

In summary, the ubiquitous presence of bursty star formation in low-mass/high-$z$ galaxies predicted by the state-of-the-art simulations of galaxy formation calls for a careful examination of its impact on existing and forthcoming observations. By analyzing the \textit{High-Redshift} suite of FIRE-2 simulations for the observability of galaxies with bursty SFHs by \textit{JWST}, we approach this problem in the context of recent \textit{JWST} observing programs for studying galaxy formation at high redshift. We have reached the following main conclusions: 

(i) The strong time variability of the SFR in high-$z$ galaxies introduces a modulation of their brightness (and apparent size) by up to several apparent magnitudes, resulting in non-monotonic time evolution in the observability of galaxies close to the limiting depth of the survey. This effect is enhanced when the nebular emission tracing recent star formation is considered. 
The time-variable SFRs imply that, at a given $M_{\star}$, some galaxies will be detectable while others not in rest UV-selected samples. 
The stellar mass scale of $f_{\rm obs} \sim 50\%$ observability depends on survey depth and is $M_{\star} \sim 10^{8.5}$ to $10^{9}\,M_{\odot}$ at $z\sim7$ for a \textit{JWST}/NIRCam survey reaching a limiting magnitude of $m^\mathrm{lim}_\mathrm{AB} \approx 29$--30, representative of surveys such as JADES and CEERS.

(ii) Systematic offsets in galaxy properties are predicted between galaxy samples detectable and non-detectable by \textit{JWST}. This implies that non-trivial selection effects from bursty SFHs may exist in flux-limited surveys, especially for physical properties closely related to the duty cycle, such as the mass fraction of cold and dense gas and the specific star formation rate. On mass scales where galaxies are marginally detectable/non-detectable, we estimate the selection effect to cause the $f_\mathrm{gas,CD}$ to be higher by a factor of 2.5 (or 0.4\,dex) and the sSFR to be higher by up to order of 20 (or 1.2\,dex) for detectable galaxies relative to non-detectable galaxies in the same stellar mass bin. Since low-mass galaxies are the most abundant, the biases predicted for galaxies near the survey limit are representative of biases in galaxy properties averaged over entire surveys.

The observational implications of bursty star formation extend beyond what has been examined in this work. 
The assumed star formation history, if incorrectly neglecting burstiness, can greatly bias stellar masses inferred from SED modeling \citep[e.g.][]{Endsley_2022}. 
Moreover, as demonstrated in \citet{Sun_2023}, bursty SFHs can substantially affect the UV luminosity function especially at the bright end, which in turn impacts analyses based on the widely-used halo abundance matching technique \citep{Furlanetto_2022_BurstySF,Dekel_2023,Mason_2023,Munoz_2023,Shen_2023}.  
Another implication of time-variable star formation histories is that low-mass galaxies may appear quiescent between bursts of star formation \cite[][]{Looser_2023, Gelli_2023}. 
We will explore these aspects, along with effects on other key physical properties such as galaxy size and kinematics, with better statistics in future work. 
Meanwhile, the impact on observability and selection effects on inferred galaxy properties due to bursty star formation explored here are unlikely to be unique to the study of galaxies at $z>5$. Observations at lower redshift, such as at cosmic noon ($z\sim2-3$) where survey selection is also frequently done in the rest UV \citep[e.g.][]{Steidel_2014}, may also be subject to similar effects. It will be interesting for future work to characterize the effects of bursty star formation in other contexts. Finally, our analysis also motivates explorations of ways to probe bursty star formation observationally and characterize its mass and redshift dependence \cite[see e.g.][]{Munoz_2023, Sun_2023EBL}, which are essential for revealing the origin of stochasticity in the SFH of galaxies. 

\section*{Acknowledgments}
The authors thank Jordan Mirocha and Allison Strom for useful discussions. GS was supported by a CIERA Postdoctoral Fellowship. 
CAFG was supported by NSF through grants AST-2108230  and CAREER award AST-1652522; by NASA through grants 17-ATP17-0067 and 21-ATP21-0036; by STScI through grant HST-GO-16730.016-A; by CXO through grant TM2-23005X; and by the Research Corporation for Science Advancement through a Cottrell Scholar Award. The Flatiron Institute is supported by the Simons Foundation. 
The simulations used in this paper were run on XSEDE computational resources (allocations TG-AST120025, TG-AST130039, TG-AST140023, and TG-AST140064). Additional analysis was done using the Quest computing cluster at Northwestern University.

\section*{Data Availability}
The data supporting the plots and analysis in this article are available on
reasonable request to the corresponding author. 
A subset FIRE-2 simulation snapshots are publicly available at \url{http://flathub.flatironinstitute.org/fire} \citep{Wetzel_2023}. 
Additional FIRE data products are available at \url{https://fire.northwestern.edu/data}. A public version of the \textsc{Gizmo} code is available at \url{http://www.tapir.caltech.edu/~phopkins/Site/GIZMO.html}.



\bibliographystyle{mnras}
\bibliography{jwst_highz} 





\bsp	
\label{lastpage}
\end{document}